\documentclass[12pt]{article}
\usepackage[dvips]{epsfig}
\usepackage{graphicx}

\begin{document}

\title{Multiple Volume Reflection from Different Planes Inside One Bent Crystal}

\vspace{1.5truecm}
\author{
{\bf Victor Tikhomirov}\\
Research Institute for Nuclear Problems,\\
Belarus State University, Minsk, Belarus\\
tel:+375-172-264739, Fax: +375-172-265124\\
 e-mail: vvtikh@mail.ru}

\vspace{1.5truecm}

\maketitle
\vspace{1.0truecm}

\begin{abstract}
It is shown that multiple volume reflections from different planes
of one bent crystal becomes possible when particles move at a
small angle with respect to a crystal axis. Such a Multiple Volume
Reflection makes it possible to increase the particle deflection
angle inside one crystal by more than four times and can be used
to increase the efficiency of beam extraction and collimation at
the LHC and many other accelerators.
\end{abstract}
\vfill\eject

\section{Introduction}

The effect of volume reflection (VR) of charged particles from bent crystal
planes was predicted in \cite{tar} and recently observed in experiments
\cite{iv1,iv2,sca}. Its evidence was also found \cite{bir} in
collimation experiments \cite{fli,car}. Present interest to VR is
mostly inspired by the possibilities which it opens up for the beam
extraction and collimation \cite{yaz}.

A principal advantage of the VR effect is its manifestation in
the whole interval of tangential directions to bent crystal planes which
greatly simplifies the crystal alignment. However a small
value of the VR angle makes it necessary \cite{bir} to use multiple proton
passage which, still, does not allow to reduce the proton interaction rate
with crystal at the LHC considerably \cite{bir2}. The particle loss in crystal
would be considerably reduced if the VR angle was larger. In this paper we
will show that the increase of the latter is made possible
by the particle reflection from multiple sets of crystal planes
inside one bent crystal which becomes possible when protons
move at a small angle with respect to a crystal axis.

\section{Volume reflection from an inclined \\ crystal plane}

VR occurs \cite{tar} when a direction of particle motion becomes
tangential to the local direction of bent crystal planes. Till now only
the reflection from crystal planes normal to the bending plane was studied
both theoretically and experimentally \cite{tar,iv1,iv2,sca,bir,fli,car,yaz}.
In particular, an explicit expression for the angle of VR from such
"vertical" plane was obtained recently \cite{mai}.

The stationary Cartesian coordinate system $XYZ$ with $Z$ axis parallel
to a set of crystal axes at the crystal entrance and plane $XZ$ parallel
to the plane of crystal bending will be used to describe the
particle motion outside the crystal -- see Fig. 1. However to describe
the particle motion inside the latter the comoving Cartesian coordinate
system $rYz$ (see Fig. 1) is more suitable \cite{mai}. Both the longitudinal
axis $z$ and the transverse plane $rY$ of this coordinate system rotate
synchronously with the particle radius vector $\vec{r}$ by its azimuthal angle
$\varphi \simeq v t/R$, where $R$ is the bending radius, $t$ is the time equal zero
at the crystal entrance and $v$ is the particle velocity. The dynamics of the
reflection process from a "vertical" plane is described by the effective
potential $U(r)-pvr/R$, where $r$ is the radial coordinate, $U(r)$ is the planar
potential and $p$ is the particle momentum.

The comoving coordinate system is also suitable for treating the interaction
with various crystal planes touched by particles moving at a small angle with respect to a
crystal axis -- see Fig. 1.  With one exception all these planes are "inclined",
i.e. form angles $\alpha \neq \pi/2$, $\pi/2 < \alpha < \pi/2$, with the bending plane
-- see Figs 1-3. All the planes remain stationary while the horizontal velocity component $V_X-v\varphi$
(we assume that $|V_X|  \ll v$ and $\varphi  \ll 1$), on the opposite, changes with
time in the transverse plane $rY$ of the commoving coordinate system (see Fig. 2 ). As a result,
different "inclined" planes, one after another, become tangential to the particle
velocity.

However the planes are not "smooth", rather consist of axes forming the angle
\begin {equation}
\psi = \sqrt{(V_X-v\varphi)^2 + V_Y^2}
\end{equation}
with the current particle velocity. At relatively large $\psi$ the maximum particle
deflection angle by an axis is estimated by the formula
\begin {equation}
\theta_{ax} \simeq \frac{2 \pi Z e^2}{\varepsilon \psi d_{ia}} \simeq
\frac{90 Z \mu rad}{\varepsilon(TeV) \psi(\mu rad) d_{ia}(\AA) },
\end{equation}
where $\varepsilon$ is the total energy of a particle which is assumed
to be ultrarelativistic and $d_{ia}$ is an interatomic distance on the axis.
A "smooth plane" approximation in which a particle deflection by the axes
does not differ considerably from that by the continuum potential of the
plane constituted by these axes can be used if $\theta_{ax} \ll  \vartheta_c$,
where $\vartheta_c $ is the critical channeling angle of the considered
plane set. On can make certain that the latter condition is equivalent
to $\psi \gg 10 \mu rad$ in the case of $7TeV$ proton deflection by low
index $Si$ planes which we will consider as an example throughout the
paper. The predictions made using the "smooth plane" approximation will
be checked by the direct simulations of particle motion in the field of
axes in the next section. For now let us consider a particle deflection
in the planar potential of an inclined bent plane.

To evaluate the angle of VR from the latter we will introduce the rotated axes
$x$ and $y$ being, respectively, normal and parallel to the local direction
of the bent crystal plane and situated in the transverse plane $rY$ -- see Fig. 3.
The particle motion along these axes is governed, respectively, by the
effective potentials $U(r)-pvx sin\alpha/R$ and $pvy cos\alpha/R$ and
characterized by the modules
\begin {equation}
v_x(x)=\sqrt{\frac{2}{\varepsilon}\left({\varepsilon_\bot}_x-U(x)+\frac{p v}{R} x sin
\alpha \right)} \;c
\end{equation}
and
\begin {equation}
v_y(y)=\sqrt{\frac{2}{\varepsilon}\left({\varepsilon_\bot}_y-\frac{p
v}{R} y cos \alpha \right)}\; c
\end{equation}
of the corresponding velocity components, where $c$ is the speed of light.
The energies ${\varepsilon_\bot}_x$ and ${\varepsilon_\bot}_y$ of the transverse
nonrelativistic particle motion in the $x$ and $y$ directions are determined
by the coordinates $x_0, y_0$ and velocity components $v_x(x_0)$, $v_y(y_0)$
at the crystal entrance. The equation
\begin {equation}
{\varepsilon_\bot}_x-U(x_t)+\frac{p v}{R} x_t sin \alpha = 0
\end{equation}
determines the turning point x-coordinate $x_t$ at a specified
${\varepsilon_\bot}_x$ value. With a view to obtain explicit
expressions for the particle deflection angles both in $XZ$ and $YZ$
planes we will follow \cite{mai} introducing the x-component
\begin {equation}
\tilde{v}_x=\sqrt{\frac{2}{\varepsilon}\left(\varepsilon_\bot-U(x_t)+\frac{p v}{R} x sin
\alpha \right)} \; c
\end{equation}
of the velocity of particle motion in the constant plane potential $U(x_t)$
in the comoving coordinate system.

Since the turning point is the point of symmetry of the particle trajectory,
we will, as usual \cite{tar,mai}, consider only a first half of it. The auxiliary
velocity $x$-component (6) allows to write the time of particle motion from the point $x_0$,
 $x_0 > x_t$, to a point $x$, $x_t \leq x \leq x_0$, in the form
\begin {equation}
t=\int_{x_0}^{x}\frac{dx}{-v_x(x)} = \Delta t(x,R/ sin \alpha) - \frac{R}{v^2 sin \alpha}
[\tilde{v}_x(x)-\tilde{v}_x(x_0)]
\end{equation}
where
\begin {equation}
\Delta t(x,{R/sin \alpha})=\int_{x}^{x_0}\left(\frac{1}{v_x(x)}-\frac{1}{\tilde{v}_x(x)}\right)dx.
\end{equation}
Note that since $x \leq x_0$ the particle moves with negative velocity $x$-component
$-v_x(x)$ up to the turning point.

The difference of $v_x(x)$ and $\tilde{v}_x(x)$ decreases fast and,
consequently, the integral (8) saturates at $|x-x_t| > U_0 R/(p v sin \alpha)$ or at
the angles $\Delta\varphi \sim \sqrt{2 |x-x_t|/R} \gg \vartheta_c$ of the vector
$\vec{r}$ rotation. Here $U_0$ is the planar potential amplitude which determines
the critical channeling angle $\vartheta_c =\sqrt{2 U_0/p v} \simeq \sqrt{2 U_0/\varepsilon}$
value. The negative integral (8) represents itself a reduction of the time of particle
motion through the reflection region $|x-x_t| \leq U_0 R/(p v sin \alpha)$
due to the local transverse motion acceleration in the planar potential.
The time (7) which can also be expressed through the velocity y-component (4):
\begin {equation}
t=\int_{y_0}^{y}\frac{dy}{v_y(y)} = \frac{R}{v^2 cos \alpha}
[v_y(y_0)-v_y(y)]
\end{equation}
determines the angle $\varphi = v t/R \ll 1$ of the local
coordinate system rotation which enters the expression
\begin {equation}
\nonumber V_X(x) \simeq v_x(x) sin \alpha + v_y(y) cos \alpha + v_z \varphi
\end{equation}
for the particle velocity component along the stationary axis $X$. The couple of
expressions (7) and (9) allows to represent Eq. (10) in the form
\begin {eqnarray}
\nonumber V_X(x) \simeq
v_x(x) sin \alpha +v_y(y_0) cos \alpha \\+\frac{v^2 sin^2
\alpha}{R} \Delta t(x,{R/sin \alpha})- [\tilde{v}_x(x)-\tilde{v}_x(x_0)] sin \alpha.
\end{eqnarray}
Eq. (11) reduces both to the initial velocity $X$-componrent
\begin {equation}
V_X(x_0)=V_{X0} = v_x(x_0) sin \alpha  + v_y(y_0) cos \alpha
\end{equation}
at the crystal entrance and to the expression
\begin {eqnarray}
\nonumber V_X(x_t)=
\\\tilde{v}_x(x_0) sin \alpha   +v_y(y_0) cos \alpha +\frac{v^2 sin^2
\alpha}{R} \Delta t(x_t,{R/sin \alpha})
\end{eqnarray}
for the particle velocity X-component in the turning point $x_t$ in which,
according to the definitions (3), (5) and (6), one has $\tilde{v}_x(x_t) = v_x(x_t)=0$.

Applying further Eq. (13) to the intervals of particle motion before
and behind the turning point one can evaluate the reflection angle
at arbitrary value of the crystal bending angle $\varphi_b =L/R$,
where $L$ is the crystal length. However here we will consider the most
practically important case of large bending angles $\varphi_b \gg \vartheta_c$
allowing both to neglect the difference
$\tilde{v}_x(x_0)-v_x(x_0) \sim \vartheta_c^2/\varphi_b \ll \vartheta_c$ and put
$x_0 = \infty$ in Eq. (8). Doubling  then, as usual \cite{tar,mai}, the deflection
angle at the turning point, one finally obtains the particle deflection angle in the
$XZ$ plane
\begin {equation}
\theta_X=\frac{2}{v}[V_X(x_t)-V_X(x_0)]=-\theta_{R}(R/sin \alpha)sin \alpha
\end{equation}
where
\begin {equation}
\theta_{R}(R/sin \alpha)=-\frac{2 v \Delta t(x_t,R/sin \alpha)}{R/sin \alpha}=
\frac{2 v }{R/sin \alpha} \int_{x_t}^{\infty}\left(\frac{1}{\tilde{v}_x(x)}-\frac{1}{v_x(x)}\right)dx
\end{equation}
is the deflection angle in the incidence plane which is parallel both to the
$x$-axis and initial particle velocity and in which the particle trajectory is
situated. Eqs. (14) and (15), naturally, reduce to the expression \cite{mai}
for VR angle from a "vertical" bent plane at $\alpha = \pi/2$. Note also that
the angle (14) is negative for sufficiently large bending radii at any $\alpha$.

In much the same way Eqs. (7) and (9) can be used to write the velocity $Y$ component
in the form
\begin {eqnarray}
\nonumber V_Y = -v_x(x) cos  \alpha + v_y(y) sin \alpha \simeq
-v_x(x) cos \alpha + v_y(y_0) sin \alpha \\-\frac{v^2 sin
\alpha cos  \alpha}{R}  \Delta t(x,{R/sin \alpha})+
[\tilde{v}_x(x)-\tilde{v}_x(x_0)] cos \alpha
\end{eqnarray}
allowing to obtain the particle deflection angle in the $YZ$ plane
\begin {equation}
\theta_Y=\theta_{R}(R/sin \alpha) cos \alpha.
\end{equation}

Eqs. (14) and (17), naturally, describe the change of transverse velocity components
accompanying a mirror particle reflection from an inclined bent plane with the
angle of incidence equal $(\pi-\theta_R)/2$. In general, since the turning point coordinate is
determined by Eq. (5) containing the effective bending radius $R/sin \alpha$,
the angle (15) can not be reduced to the expression \cite{mai} for the angle
of reflection from a "vertical" bent plane by the replacement of the
integration variable. Such a reduction, fortunately, becomes possible at
$R \gg U_0/(pvd)$, where $d$ is an inter-planar distance, when the angle
(15) approaches the limit for infinite bending radius which does not depend
on both $R$ and $\alpha$. In other words, the reflection angle (15) is bound
from above by the same value of about $1.6 \vartheta_c$ \cite{tar} as that
from "vertical" plane.

\section{MVR inside a crystal}

The angles (14) and (17) of particle deflection from an inclined plane allow to
evaluate the total angle of MVR from multiple crystal plane sets touched
by a particle inside a bent crystal. An important point here is that inclined planes
constitute symmetrical pairs with complimentary inclination angles $\alpha$ and
$\pi-\alpha$ -- see Figs. 1 and 2. The reflection angles (14) in the $XZ$ plane
from such symmetrical crystal planes sum up opening a possibility to considerably
increase the angle of the VR in the bending plane $XZ$. On the other hand, the angles
(17) of reflection in the vertical $YZ$ plane are opposite in sign (see Fig. 2), allowing
the angle of VR in this plane to vary from several critical channeling
angles down to zero. Note that significant particle deflection either in the
process of MVR inside a crystal or at repetitive passages through it will
violate the symmetry of particle motion in the bending plane and can cause
both "switching on" and "off" of some crystal plane sets from the
MVR process. As a consequence the latter will start giving its contribution to the
particle deflection in the $YZ$ plane. Thus, to maintain the horizontal direction
of MVR a special care in choosing crystal length, bending angle and orientation
in the horizontal plane should be taken. To simplify the situation we will
make here the angle of MVR in the vertical plane equal to zero restricting our
consideration to the case of MVR in the crystal bending plane $XZ$.

Let as consider a multiple reflection of $7 TeV$ protons from the bent
crystal planes passing through the $<111> Si$ axis as an example. One of
the three strongest planes $(1\bar{1}0)$ passing through this axis will be
directed at a right angle to the bending plane playing the role of "vertical"
plane of "single" VR -- see Figs. 1 and 2. We will also hold fixed
both the crystal bending angle $\varphi_b = L/R = 100 \mu rad$ and
symmetric direction $V_{X0}/v=0.5 \varphi_b =50 \mu rad$ of particle
incidence in the $XZ$ plane (see Fig. 2) and investigate the
dependence of the MVR angle on the bending radius varying the latter
proportionally to the crystal length $L$.

In order to give its contribution to MVR, a crystal plane has
to have an inclination angle $\alpha$ exceeding the angle
$\alpha_{inc}=arctg V_{Y0}/V_{X0}$ of particle
incidence with respect to the bending plane $XZ$ (see Fig. 2). This
condition considerably limits the initial Y-component of particle velocity.
In order to use the rest two strongest planes $(\bar{1}01)$ and $(0\bar{1}1)$
having inclination angles $\alpha=30^\circ$ and $\alpha'=\pi-\alpha=150^\circ$
we will choose the value $V_{Y0}=25  \mu rad$ corresponding to
$\alpha_{inc} \simeq 27^\circ < \alpha$. The main planes $(\bar{1}01), (\bar{3}12),
(\bar{2}11), (\bar{3}21)$, $(1\bar{1}0)$, $(2\bar{3}1)$, $(1\bar{2}1)$, $(1\bar{3}2)$ and
$(0\bar{1}1)$ will give their contributions to MVR at such an $\alpha_{inc}$
choice, while the planes $(\bar{2}\bar{1}3)$, $(\bar{1}\bar{2}3)$ and $(\bar{1}2\bar{1})$
will not -- see Fig. 2. Note that the chosen $V_{Y0}$ value makes the "smooth plane"
approximation applicable, at least for the low index crystal planes.

The bending radius dependence of contributions of the low-index planes to the
MVR angle in the $XZ$ plane is illustrated by Fig. 4. The bending radius is measured
in units of $R_{min}=11.34 m$. Fig. 4 indeed demonstrates that an additional reflection from
the inclined planes increases the total particle deflection angle about four times.

To confirm this prediction based, in fact, on the "smooth plane" approximation direct
simulations of the $7 TeV$ proton motion in the field of $<111> Si$ axes should be
used. Our simulation procedure, in many respects reminding that described
in \cite{akh}, was considerably accelerated by the use of the approximation
\begin{equation}
V(\rho) = V_0 \times \cases{ 1 - \rho^2/3 \rho^2_1,
\hspace{12mm} 0 \leq \rho \leq \rho_1;  \cr 2 \rho_1/3
\rho, \hspace{18mm} \rho_1 \leq \rho \leq \rho_{max};  \cr 0,
\hspace{21mm} \rho \geq \rho_{max}}
\end{equation}
to the potential of the $<111>$ $Si$ axis in which $\rho_1 \simeq 0.14 \AA$, $V_0 \simeq 107 eV$
and $\rho_{max} = a/2\sqrt{6} \simeq 1.109\AA$, where $a \simeq 5.431 \AA$
is the $Si$ lattice constant. This approximation allows to treat
particle motion analytically both in the field of each axis and in between the
neighboring $<111>$ axes constituting a two-dimensional hexagonal lattice in the transverse
plane (see Fig. 1). These semi-analytical simulations allow to obtain such
two-dimensional particle distributions in scattering angles as that shown in Fig. 5
for $R=100m$ and $L=1cm$.

Corresponding one-dimensional distribution in $\theta_Y$ turns out
to be nearly symmetric, demonstrating the absence of large coherent deflection
effects in the $YZ$ plane predetermined by the opposite particle deflections
by the paired symmetrical planes. The same situation is observed in a wide region of
$V_{Y0}$ variation at symmetric particle incidence with $V_{X0}=v \varphi_b/2$
in the bending plane. At some $V_{Y0}$ values, however, a capture into the
channeling regime motion in the field of various crystal planes at the crystal
entrance makes the distribution in $\theta_Y$ angle strongly asymmetric even
at $V_{X0}=v \varphi_b/2$. In general, particle capture by the numerous
crystal planes at the crystal entrance, becoming possible when particles
move at small angles with respect to a crystal axis, opens up other new
possibilities for beam extraction and collimation which will be described elsewhere.

The distributions in $\theta_X$ angle present in Fig. 6 for several crystal
bending radii confirm in general the predictions obtained in the "smooth
plane" approximation. The positions of maxima of these and several
other distributions are shown by circles in Fig. 4 along with an interpolating
curve. One can see that the "smooth plane" approximation slightly overestimates
the MVR angle at small and considerably underestimates it at large bending radii.
In principle, the particle scattering from all the high-index planes can be
additionally taken into consideration in the "smooth plane" approximation to fill
the gap between the two upper curves at large bending radii. Some of such planes,
for instance, $(\bar{4}13), (\bar{5}14)$ and $(\bar{6}15)$, act coherently leading
to the cumulative deflection angle distinguishable in Fig. 2. The "smooth plane"
approximation is much less applicable for these planes than for the low-index ones
at the relatively small $V_{Y0}$ values necessary to involve more crystal planes in
the MVR process in relatively short crystals. In addition, a shallow planar
potential of high-index planes leads to relatively high particle volume caption
probability. All these complications make a direct simulation of the particle
motion in the field of crystal axes the preferred method of the MVR study.
The "smooth plane" approximation, nevertheless, gives both the simplest
picture and correct intuitive estimate of the Multiple Volume Reflection effect,
which can be applied for both other crystal elements and orientations as well as
for the negative particle case.

For the purpose of comparison, the distribution of protons reflected from a single
vertical plane bent with $200 m$ radius is also shown in Fig. 6 on the right.
Figs. 4-6 visually demonstrate that the angles of MVR can more than
four times exceed that of the "single" VR from a "vertical" bent plane.
One can also see from Fig. 4 that the angular width of the MVR peaks
exceeds that of the "single" VR one.

The MVR distributions are also more asymmetric than the "single" VR one. At
this 10\% - 15\% of the protons are deflected in the direction of crystal
bending acquiring a positive velocity X-component (see Figs. 5 and 6).
At "single" VR such protons amounts to 3-5\% appearing because of the volume
capture into the channeling regime motion due to the incoherent scattering
by both electrons and nuclei. However, when particles move at small angles
with respect to crystal axes constituting crystal planes, additional volume
capture becomes possible due to the coherent particle scattering by separate crystal axes.
Note that we have not considered the incoherent particle scattering
by electrons which has to increase the particle volume caption additionally. On the
one hand, the volume caption and deflection in the direction of crystal bending
of ten or fifteen percent of the particles will make it more difficult to deflect
the most part of them by a large angle using MVR in a long sequence
of bent crystals. On the other, the increase of the mean reflection angle, of its
angular dispersion and of the number of volume captured particles all add up to
the broadening of the beam angular distribution. As a result, the mean square
deflection angle of all the incident particles, important for efficient beam
collimation \cite{yaz,bir2}, can exceed $10 \mu rad$ in a 1 cm $Si$ crystal
in the MVR conditions.

\section{Conclusions}

The main idea of the paper was that particles moving at small angles with respect
to a crystal axis in a bent crystal experience volume reflection from the
numerous crystal planes passing through this axis. Both the newly developed model
of particle reflection from inclined crystal planes and the direct simulations of particle
motion in the field of crystal axes demonstrate that the total MVR angle exceeds
that of the VR from a "single" "vertical" plane more than four times. The particle
deflection angle magnification by the MVR, also used in series of parallel
strip-like crystals, multiple-strip crystals, crystals periodically curved by groves
or alternative tensile coating opens up new possibilities to improve the efficiency of both
the beam deflection and collimation at the LHC and many other accelerators.

{\sl Acknowledgement}

The author gratefully acknowledges useful discussions with V.G.
Baryshevsky, X. Artru,  V.A. Maisheev and I.A. Yazynin. This work was partly
supported by the \# 03-52-6155 INTAS Project.

\vfill\eject

\newpage
\begin{figure}[!ht]
\centering \psfull 
    \epsfig{file=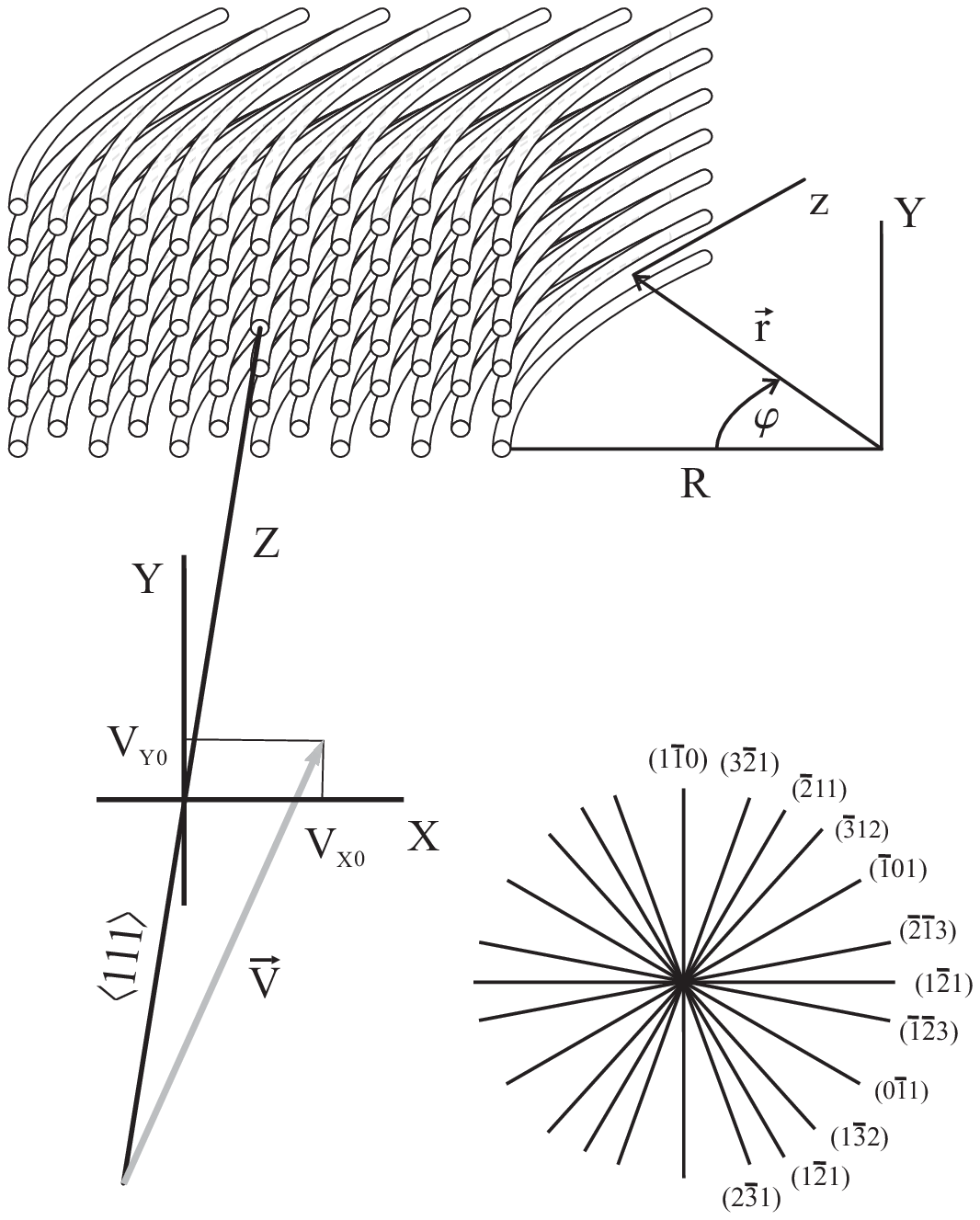,width=10cm}
\vspace{30mm}
    \caption{Particles hit a crystal moving at a small angle with respect
to a set of crystal axes. $XYZ$ and $xyz$ are, respectively, stationary and comoving
coordinate systems. Main crystal planes passing through the $<111>$ $Si$
axis are also shown.}
\end{figure}

\newpage
\begin{figure}[!ht]
\centering \psfull 
    \epsfig{file=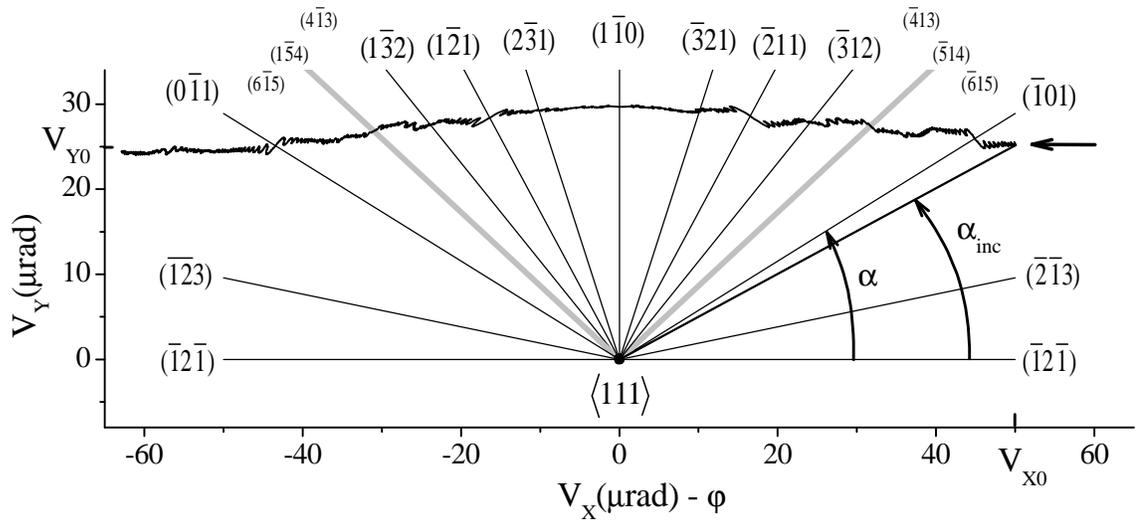,width=15cm}
    \caption{Particle velocity evolution and stationary crystal plane directions in
the transverse $rY$ plane of the comoving coordinate system. Particle velocity becomes tangential
to different sets of crystal planes parallel to $<111>$ axis.}
\end{figure}

\vspace{30mm}

\begin{figure}[!ht]
\centering \psfull 
    \epsfig{file=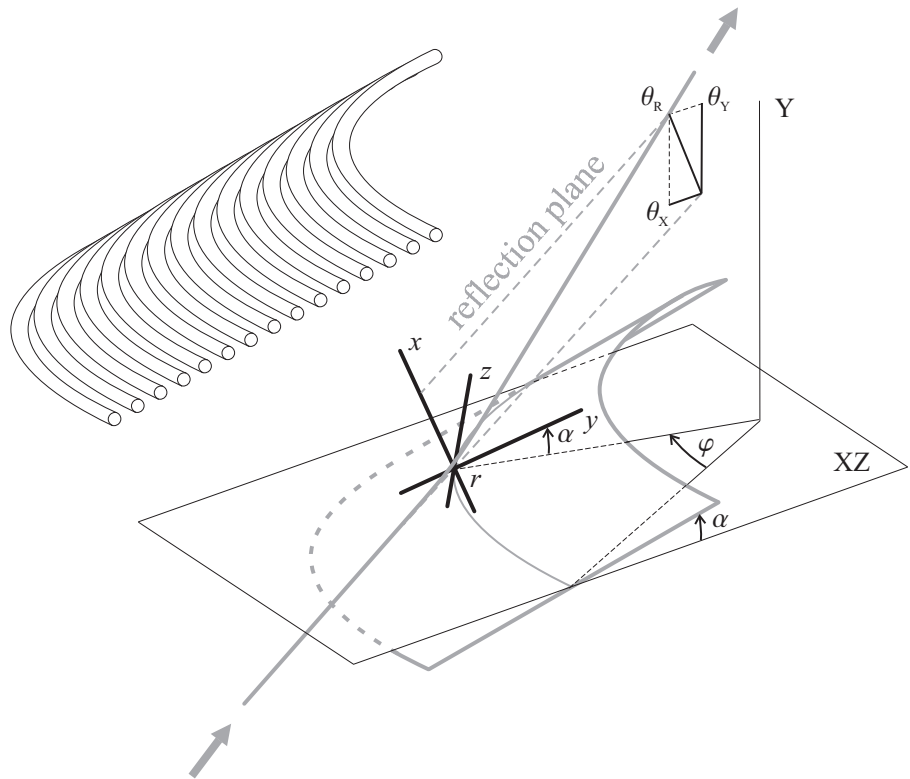,width=12cm}
    \caption{Comoving coordinate system $xyz$ used to describe the VR from
an "inclined" plane constituted by crystal axes. Both $x$ and $y$ axes are situated
in the $rY$ plane.}
\end{figure}

\newpage
\begin{figure}[!ht]
\centering \psfull \vspace{-10mm}
    \epsfig{file=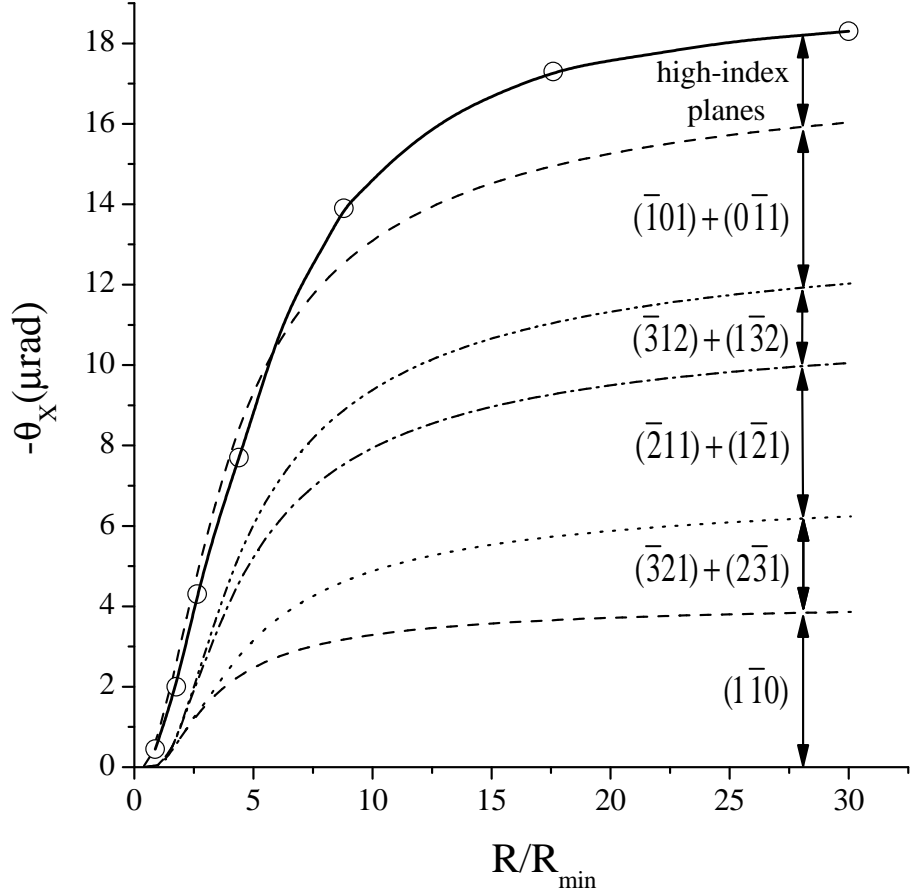,width=12cm}
    \caption{Dependence of contributions of the low index $Si$ planes to the
angle of $7 TeV$ proton MVR on the crystal bending radius. Circles with interpolating
curve represent the results of Monte Carlo simulations of proton motion in the
potential of bent axes.}
\end{figure}

\newpage
\begin{figure}[!ht]
\centering \psfull 
    \epsfig{file=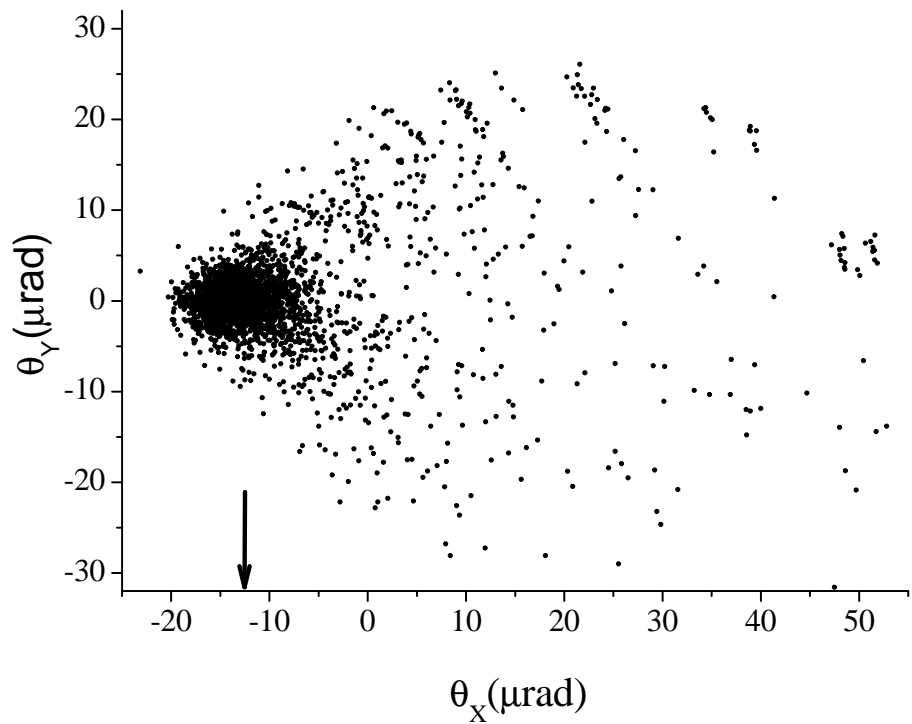,width=12cm}
    \caption{Distribution in scattering angles of 7 TeV protons after passage through a
1 cm Si crystal with bending radius $R=100m$.}
\end{figure}

\newpage
\begin{figure}[!ht]
\centering \psfull 
    \epsfig{file=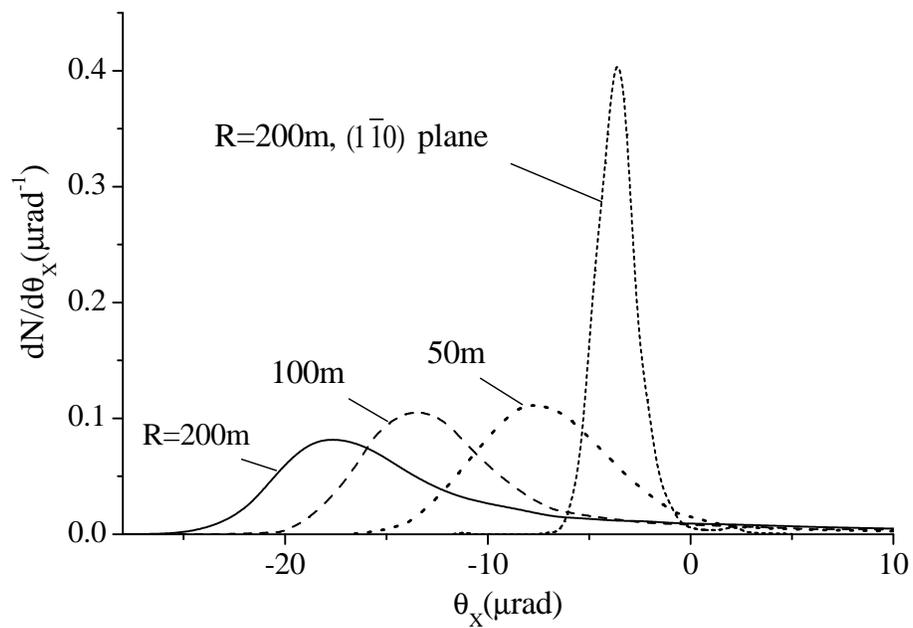,width=12cm}
    \caption{Angular distribution of 7 TeV protons in the $XZ$ plane of $Si$ crystals
with indicated bending radii. The right curve illustrates "single" VR from $(1\bar{1}0)$
plane.}
\end{figure}

\end{document}